# Multi-Grid Monte Carlo
# IV. One-Dimensional $O(4)$-Symmetric Nonlinear $\sigma$-Model

Tereza Mendes
Alan D. Sokal
*Department of Physics*
*New York University*
*4 Washington Place*
*New York, NY 10003 USA*
Internet: `MENDES@MAFALDA.PHYSICS.NYU.EDU`, `SOKAL@ACF4.NYU.EDU`

March 27, 1995

**Abstract**

We study the dynamic critical behavior of the multi-grid Monte Carlo (MGMC) algorithm with piecewise-constant interpolation and a W-cycle, applied to the one-dimensional $O(4)$-symmetric nonlinear $\sigma$-model [= $SU(2)$ principal chiral model], on lattices from $L = 128$ to $L = 16384$. Our data for the integrated autocorrelation time $\tau_{int,\mathcal{M}^2}$ are well fit by a logarithmic growth. We have no idea why the critical slowing-down is not completely eliminated.

By now it is widely recognized [1,2,3,4] that better simulation algorithms, with strongly reduced critical slowing-down, are needed for high-precision Monte Carlo studies of statistical-mechanical systems near critical points and of quantum field theories (such as QCD) near the continuum limit. One promising class of such algorithms is *multi-grid Monte Carlo* (MGMC) [5,6,7,8].

In the first paper of this series [6], we explained the conceptual foundations of the MGMC method, and proved the absence of critical slowing-down for Gaussian (free-field) MGMC with piecewise-constant interpolation and a W-cycle. We[1] conjectured [6, Section IX.C] that for asymptotically free theories — which are "almost Gaussian" except at very long distances — the critical slowing-down should likewise be completely eliminated except for a possible logarithm. However, when we made a numerical study of MGMC in the two-dimensional $O(N)$-symmetric nonlinear $\sigma$-model (also called $N$-vector model) with $N = 4$ [8], we found to our surprise that the dynamic critical exponent is *not* zero: $z_{int,\mathcal{M}^2} = 0.60 \pm 0.07$ (subjective 68% confidence interval). We then produced [8, Section 4.1] a heuristic explanation of this fact, based on the logarithmically decaying deviations from Gaussianness in an asymptotically free theory. This explanation does not, unfortunately, give a quantitative prediction for $z_{MGMC}$; it merely makes the weak prediction that $0 < z_{MGMC} < z_{heat-bath}$.

Richard Brower (private communication) then pointed out to us that if our explanation is correct, it follows that for a *one*-dimensional $\sigma$-model — in which the deviations from Gaussianness decay much more rapidly, namely as a power law — the critical slowing-down should be *completely* eliminated, i.e. we should have $z_{MGMC} = 0$ (with *no* logarithm). We hardly doubted that this would be the case, but we decided to make an empirical test to settle the matter once and for all.

In this paper we report evidence which suggests that in fact the critical slowing-down is *not* completely eliminated in the one-dimensional $O(4)$-symmetric $\sigma$-model. More precisely, we have simulated this model on periodic lattices of size $L = 128$ through $L = 16384$, using MGMC with piecewise-constant interpolation and a W-cycle, and we find that the autocorrelation time is well fit by a logarithmic growth $\tau_{int,\mathcal{M}^2} \approx A + B \log L$. (For $L \gtrsim 512$ the data can also be well fit by a power-law growth with exponent $z_{int,\mathcal{M}^2} \approx 0.18 \pm 0.01$, but the logarithmic fit is better.) Of course, it is perfectly possible that the autocorrelation time will level off at larger lattice sizes — even at $L = 16384$ it has only reached $\approx 5$ sweeps — but the data for $L \leq 16384$ show no tendency in this direction. We are thus left with the distinct impression that the critical slowing-down is *not* completely eliminated in this model — even if only a logarithm remains — and we have no idea why.

The model in question is defined by the Hamiltonian (= Euclidean action)

$$H(\boldsymbol{\sigma}) = -\beta_{O(4)} \sum_{x=1}^{L} \boldsymbol{\sigma}_x \cdot \boldsymbol{\sigma}_{x+1} , \tag{1}$$

where each spin $\boldsymbol{\sigma}_x$ is a unit vector in $\mathbb{R}^4$, and $\beta_{O(4)} > 0$ is the inverse temperature in the $O(4)$ normalization; periodic boundary conditions ($\boldsymbol{\sigma}_{L+1} = \boldsymbol{\sigma}_1$) are employed. Since the

---

[1] Here "we" refers to the second author only. The first author is not responsible for the second author's sordid past.

sphere $S^3$ is isometric (as a Riemannian manifold) to the group $SU(2)$ via the map

$$g = \begin{pmatrix} \sigma^0 + i\sigma^3 & \sigma^2 + i\sigma^1 \\ -\sigma^2 + i\sigma^1 & \sigma^0 - i\sigma^3 \end{pmatrix}, \tag{2}$$

this model can also be thought of as the $SU(2)$ principal chiral model

$$H(g) = -\beta_{SU(2)} \sum_{x=1}^{L} \operatorname{Re} \operatorname{tr}(g_x^\dagger g_{x+1}), \tag{3}$$

with $\beta_{SU(2)} = \beta_{O(4)}/2$. Our MGMC algorithm employs this $SU(2)$ representation, and is described in detail in [8]. We used a W-cycle with one heat-bath pre-sweep and one heat-bath post-sweep ($m_1 = m_2 = 1$); the coarsest grid always has two sites. All runs used a random initial configuration ("hot start").

We measured static quantities (expectations) and dynamic quantities (autocorrelation functions) for the following observables:

$$\mathcal{M}^2 = \left(\sum_x \boldsymbol{\sigma}_x\right)^2 \tag{4}$$

$$\mathcal{F} = \left|\sum_x e^{2\pi i x/L} \boldsymbol{\sigma}_x\right|^2 \tag{5}$$

$$\mathcal{E} = \sum_x \boldsymbol{\sigma}_x \cdot \boldsymbol{\sigma}_{x+1} \tag{6}$$

The mean values of these observables give information on different aspects of the 2-point function

$$G(x) = \langle \boldsymbol{\sigma}_0 \cdot \boldsymbol{\sigma}_x \rangle \tag{7a}$$

$$\tilde{G}(p) = \sum_x e^{ip \cdot x} \langle \boldsymbol{\sigma}_0 \cdot \boldsymbol{\sigma}_x \rangle \tag{7b}$$

In particular, we are interested in the *susceptibility*

$$\chi = \tilde{G}(0) = L^{-1} \langle \mathcal{M}^2 \rangle \tag{8}$$

and the analogous quantity at the smallest nonzero momentum

$$F = \tilde{G}(p)|_{|p|=2\pi/L} = L^{-1} \langle \mathcal{F} \rangle. \tag{9}$$

By combining these we can obtain the *second-moment correlation length*

$$\xi = \left(\frac{(\chi/F) - 1}{4\sin^2(\pi/L)}\right)^{1/2}. \tag{10}$$

The dynamic quantities for an observable $A$ ($= \mathcal{M}^2$, $\mathcal{F}$ or $\mathcal{E}$) are defined from the *unnormalized autocorrelation function*

$$C_A(t) = \langle A_s A_{s+t} \rangle - \langle A \rangle^2, \tag{11}$$

where expectations are taken *in equilibrium*, and the corresponding *normalized autocorrelation function*

$$\rho_A(t) = C_A(t)/C_A(0) \,. \tag{12}$$

We then define the *integrated autocorrelation time*

$$\begin{aligned}\tau_{int,A} &= \frac{1}{2} \sum_{t=-\infty}^{\infty} \rho_A(t) \\ &= \frac{1}{2} + \sum_{t=1}^{\infty} \rho_A(t) \,. \end{aligned} \tag{13}$$

[The factor of $\frac{1}{2}$ is purely a matter of convention; it is inserted so that $\tau_{int,A} \approx \tau$ if $\rho_A(t) \approx e^{-|t|/\tau}$ with $\tau \gg 1$.] We also define the *exponential autocorrelation time*

$$\tau_{exp,A} = \limsup_{|t| \to \infty} \frac{|t|}{-\log|\rho_A(t)|} \,. \tag{14}$$

The integrated autocorrelation time controls the statistical error in Monte Carlo measurements of $\langle A \rangle$, while the exponential autocorrelation time measures the "slowest mode" in the system (provided that it is not orthogonal to $A$). We estimated the integrated autocorrelation time by standard procedures of statistical time-series analysis [9,10], using a self-consistent truncation window of width $c\tau_{int,A}$ with $c = 8$: see [1, Section 3] and [11, Appendix C]. The choice $c = 8$ is reasonable whenever the autocorrelation function $\rho_A(t)$ decays roughly exponentially, a behavior that we will confirm explicitly here. We estimated the exponential autocorrelation time by the rather subjective procedure of looking for a plateau in the estimates of $1/\log[\rho_A(t)/\rho_A(t+1)]$ as a function of $t$; the error bars are also subjective. Of course, all estimates of $\tau_{exp,A}$ are really only approximate *lower bounds* on $\tau_{exp,A}$; it is impossible to rule out the existence of a (very-)slowly-decaying component of (very) small amplitude.

The results of our runs are shown in Table 1. The exact finite-volume values of $\chi$ and $\xi$ are obtained from [12]. The deviation from the exact values has a chi-squared of 27.75 (28 DF, confidence level = 47.8%) for the susceptibility, and 31.58 (28 DF, confidence level = 29.2%) for the correlation length.[2] The energy $E$ is not shown here, but it also agrees with the exact values (chi-squared = 24.12, 28 DF, confidence level = 67.5%). The autocorrelation time of the energy, $\tau_{int,\mathcal{E}}$, is uniformly less than 0.6, so we do not bother to report it.

In Figure 1 we plot $\tau_{int,\mathcal{M}^2}$ versus $\xi(L)/L$, indicating the different lattice sizes by symbols. Clearly $\tau_{int,\mathcal{M}^2}$ is an increasing function of $L$ at each fixed $\xi(L)/L$, and there is no evidence that the increase is slowing down as $L$ grows.

In Figure 2 we show a log-log plot of $\tau_{int,\mathcal{M}^2}$ versus $L$ for the data at $\beta/L = 40/128 = 0.3125$ (corresponding to $\xi(L)/L \approx 0.2$). In Table 2 we report the results of a weighted least-squares fit to $\tau_{int,\mathcal{M}^2} = A L^{z_{int,\mathcal{M}^2}}$ for these data, using lattice sizes $L \geq L_{min}$ and trying various values of $L_{min}$. For $L_{min} = 128, 256$ the estimates are clearly afflicted with corrections to scaling: this can be seen in the chi-squared values, as well as in the clear trend in the estimates of $z_{int,\mathcal{M}^2}$ and $A$; it is also apparent in the plot. For $L_{min} = 512$ there *may* be corrections to scaling on the borderline of statistical significance: the point at $L = 512$ is about $2\sigma$ below the trend-line of the larger lattices, with a deviation of the same sign as for $L = 128, 256$. Being conservative, we therefore take $L_{min} = 1024$ as our preferred fit: this yields the estimate $z_{int,\mathcal{M}^2} = 0.174 \pm 0.011$ (68% confidence interval). Less conservative readers might take $L_{min} = 512$, yielding $z_{int,\mathcal{M}^2} = 0.184 \pm 0.008$.

---

[2]Confidence level is the probability that the chi-squared would equal or exceed the observed value, assuming that the underlying statistical model ("null hypothesis") is correct.

This very low (though nonzero) estimate for $z_{int,\mathcal{M}^2}$ suggests that the true behavior might be logarithmic. This suggestion is supported by the fact that in Figure 1 the points at each fixed $\xi(L)/L$ are roughly equally spaced. In Figure 3 we show a semi-log plot of $\tau_{int,\mathcal{M}^2}$ versus $L$ for the data at $\beta/L = 0.3125$, and in Table 3 we show the corresponding weighted least-squares fit to $\tau_{int,\mathcal{M}^2} = A + B \log L$. The quality of the fit is much better than for the pure power-law Ansatz: even for $L_{min} = 128$ the chi-squared is extremely small, and we obtain the estimates $A = -1.456 \pm 0.090$, $B = 0.6623 \pm 0.0145$ (68% confidence intervals). We are left with the distinct impression that $\tau_{int,\mathcal{M}^2}$ indeed grows logarithmically. Of course, a definitive test of this conjecture would require lattices considerably larger than those used here (say, up to $L \sim 10^6$). Examination of Figure 1 suggests that a similar logarithmic growth holds at each fixed $\xi(L)/L$, but the constants $A$ and $B$ clearly depend on $\xi(L)/L$.

Next we test the dynamic finite-size-scaling Ansatz[3]

$$\tau_{int,A}(\beta, L) \approx \xi(\beta, L)^{z_{int,A}} g_A\big(\xi(\beta, L)/L\big) \tag{15}$$

for $A = \mathcal{M}^2$. Here $g_A$ is an unknown scaling function, and $g_A(0) \equiv \lim_{x \downarrow 0} g_A(x)$ is supposed to be finite and nonzero.[4] We impose $z_{int,\mathcal{M}^2} = 0.174$ — which is a reasonable "effective exponent" for our range of $L$ — and plot $\tau_{int,\mathcal{M}^2} \xi^{-z_{int,\mathcal{M}^2}}$ versus $\xi(L)/L$; the points should fall on a single curve, at least asymptotically for large $L$. The result is shown in Figure 4. There are small corrections to scaling for $L \leq 512$, just as observed in Table 2; but the points do clearly appear to be approaching a limiting curve.

Finally, we want to test the more detailed dynamic finite-size-scaling Ansatz

$$\rho_A(t; \beta, L) \approx |t|^{-p_A} h_A\big(t/\tau_{exp,A}(\beta, L); \xi(\beta, L)/L\big), \tag{16}$$

---

[3] Our preceding results suggest that we ought to use instead a logarithmic Ansatz, e.g.

$$\tau_{int,A}(\beta, L) \approx [\log \xi(\beta, L) + c(\xi(\beta, L)/L)] g_A\big(\xi(\beta, L)/L\big).$$

But it is difficult to know what to use for the function $c(\xi(\beta, L)/L)$. If one simply uses $c \equiv 0$, the agreement is poor; clearly the additive constant cannot be neglected for our range of $L$, as is obvious already from Figures 1 and 3 and Table 3. So we decided to use instead the power-law Ansatz as a reasonable "phenomenological" fit.

[4] It is of course equivalent to use the Ansatz

$$\tau_{int,A}(\beta, L) \sim L^{z_{int,A}} \widehat{g}_A\big(\xi(\beta, L)/L\big),$$

and indeed the two Ansätze are related by $\widehat{g}_A(x) = x^{z_{int,A}} g_A(x)$. However, to determine whether $\lim_{x \downarrow 0} g_A(x) = \lim_{x \downarrow 0} x^{-z_{int,A}} \widehat{g}_A(x)$ is nonzero, it is more convenient to inspect a graph of $g_A$ than one of $\widehat{g}_A$.

where $p_A$ is an unknown exponent and $h_A$ is an unknown scaling function. Summing (16) over $t$, it follows that[5]

$$\tau_{int,A} \sim \tau_{exp,A}^{1-p_A} \,. \tag{17}$$

Therefore (16) can equivalently be rewritten as

$$\rho_A(t;\beta,L) \approx |t|^{-p_A} \hat{h}_A\left(t/\tau_{int,A}(\beta,L)^{1/(1-p_A)}\,;\, \xi(\beta,L)/L\right) \,. \tag{18}$$

In Figure 5(a) we test this latter Ansatz for $A = \mathcal{M}^2$, using the inspired guess $p_{\mathcal{M}^2} = 0$, for the data at $\beta/L = 0.3125$. The data fall beautifully on a single curve (until the signal is obscured by the statistical noise), in excellent agreement with (18).

Thus, although $p_A = 0$ does not hold in general, it does appear empirically to hold (at least for the global observable $A = \mathcal{M}^2$) in our one-dimensional model. In such a situation, the ratio $\tau_{int,\mathcal{M}^2}/\tau_{exp,\mathcal{M}^2}$ tends as $L \to \infty$ to a limiting value (which could depend on $\xi(L)/L$) that is characteristic of the dynamic universality class. We can estimate this value by a linear fit

$$\log \rho_{\mathcal{M}^2}(t) = a - bt/\tau_{int,\mathcal{M}^2} \tag{19}$$

to the data in Figure 5(a), restricting attention to the interval $1.5 \le t/\tau_{int,\mathcal{M}^2} \le 3$ where the plot indeed appears linear; the slope $b$ gives the ratio $\tau_{int,\mathcal{M}^2}/\tau_{exp,\mathcal{M}^2}$. We obtain $a = -0.18$, $b = 0.90$. Of course, as noted above, all estimates of $\tau_{exp,\mathcal{M}^2}$ are really only approximate *lower bounds*; so the true ratio $\tau_{int,\mathcal{M}^2}/\tau_{exp,\mathcal{M}^2}$ might be lower than this.

In Figure 5(b) we make an identical plot, but now using the data from *all* values of $\beta/L \ge 0.3125$. (The data at smaller $\beta/L$ are omitted because their statistics are much poorer.) Surprisingly, the points continue to fall well on a single curve; the linear fit (19) on the interval $1.5 \le t/\tau_{int,\mathcal{M}^2} \le 3$ yields $a = -0.16$, $b = 0.92$. This indicates that the dependence on $\xi(L)/L$ in (16)/(18), if any, is extremely weak.

We also attempted to estimate $\tau_{exp,\mathcal{M}^2}$ by looking for a plateau in the estimates of $1/\log[\rho_{\mathcal{M}^2}(t)/\rho_{\mathcal{M}^2}(t+1)]$ as a function of $t$. More specifically, we used the largest $t$ before the values of $1/\log[\rho_{\mathcal{M}^2}(t)/\rho_{\mathcal{M}^2}(t+1)]$ begin to decrease or oscillate significantly. This roughly corresponds to choosing $t$ by any one of the following three criteria[6]:

(a) $t \approx 1.7\tau_{int,\mathcal{M}^2}$

(b) $t$ such that $\rho_{\mathcal{M}^2}(t) \approx 0.17$

(c) $t$ such that $\rho_{\mathcal{M}^2}(t) \approx 50[\tau_{int,\mathcal{M}^2}/(\text{Sweeps} - \text{Discard})]^{1/2}$

The error bars are subjective 68% confidence intervals. The resulting estimates of $\tau_{exp,\mathcal{M}^2}$, for the data at $\beta/L = 0.3125$, are given in Table 4. We see that the ratio $\tau_{int,\mathcal{M}^2}/\tau_{exp,\mathcal{M}^2}$ is essentially independent of $L$ and equals $\approx 0.90$. This agrees with the estimate from Figure 5(a). Again, all estimates of $\tau_{exp,\mathcal{M}^2}$ are only approximate lower bounds.

---

[5]Note that, contrary to much belief, $\tau_{int,A}$ and $\tau_{exp,A}$ do *not* in general have the same dynamic critical exponent $z$; rather, we have from (17) the dynamic scaling law $z_{int,A} = (1-p_A)z_{exp,A}$. Thus, $z_{int,A} = z_{exp,A}$ only if $p_A = 0$. For further discussion, see [1,4].

[6]All three criteria translate, in different ways, the intuitive idea of using the largest $t$ such that the signal-to-noise ratio is "not too small". In particular, criterion (c) comes from the fact that the statistical error bar on $\rho_A(t)$ is roughly of order $[\tau_{int,A}/(\text{Sweeps} - \text{Discard})]^{1/2}$: see [1,9,10,11].

We wish to thank Rich Brower, Attilio Cucchieri and Andrea Pelissetto for helpful discussions. These computations were carried out on the Convex C-210† at the NYU Academic Computing Facility and on the Cray C-90 at the Pittsburgh Supercomputing Center (PSC). The authors' research was supported in part by NSF grant DMS-9200719 and PSC grants PHY890035P and MCA94P032P.

| | | | | | | | | |
|---|---|---|---|---|---|---|---|---|
| \multicolumn{9}{c}{$d=1$ $O(4)$ model} | | | | | | | | |
| $L$ | $\beta$ | Sweeps | Discard | $\chi$ | $\chi_{exact}$ | $\xi$ | $\xi_{exact}$ | $\tau_{int,\mathcal{M}^2}$ |
| 128 | 10 | 50000 | 5000 | 12.74 ( 0.04) | 12.72 | 6.41 ( 0.15) | 6.34 | 0.59 (0.01) |
| 128 | 20 | 42869 | 5000 | 26.03 ( 0.10) | 26.02 | 13.05 ( 0.13) | 13.00 | 0.93 (0.03) |
| 128 | 40 | 200000 | 5000 | 51.41 ( 0.08) | 51.41 | 26.13 ( 0.07) | 26.14 | 1.75 (0.03) |
| 128 | 60 | 200000 | 5000 | 70.51 ( 0.08) | 70.56 | 37.94 ( 0.08) | 37.96 | 1.97 (0.04) |
| 128 | 80 | 200000 | 5000 | 82.93 ( 0.06) | 82.89 | 47.84 ( 0.09) | 47.78 | 1.81 (0.03) |
| 128 | 100 | 200000 | 5000 | 90.99 ( 0.05) | 90.98 | 56.02 ( 0.09) | 56.02 | 1.65 (0.03) |
| 128 | 120 | 200000 | 5000 | 96.64 ( 0.04) | 96.63 | 63.24 ( 0.09) | 63.21 | 1.58 (0.03) |
| 256 | 20 | 50000 | 5000 | 25.86 ( 0.09) | 26.02 | 12.42 ( 0.32) | 13.00 | 0.70 (0.02) |
| 256 | 40 | 50000 | 5000 | 52.60 ( 0.22) | 52.67 | 26.28 ( 0.27) | 26.33 | 1.23 (0.04) |
| 256 | 80 | 200000 | 5000 | 103.30 ( 0.19) | 103.38 | 52.56 ( 0.16) | 52.60 | 2.21 (0.04) |
| 256 | 120 | 200000 | 5000 | 141.54 ( 0.17) | 141.48 | 76.26 ( 0.18) | 76.18 | 2.31 (0.04) |
| 256 | 160 | 200000 | 5000 | 166.04 ( 0.14) | 166.02 | 95.84 ( 0.19) | 95.78 | 2.17 (0.04) |
| 256 | 200 | 200000 | 5000 | 182.22 ( 0.11) | 182.12 | 112.39 ( 0.20) | 112.24 | 2.06 (0.04) |
| 256 | 240 | 200000 | 5000 | 193.17 ( 0.10) | 193.37 | 126.11 ( 0.21) | 126.58 | 1.98 (0.04) |
| 512 | 40 | 31080 | 5000 | 52.78 ( 0.27) | 52.68 | 27.07 ( 0.84) | 26.33 | 0.82 (0.03) |
| 512 | 80 | 31159 | 5000 | 106.51 ( 0.61) | 105.98 | 53.85 ( 0.75) | 53.00 | 1.42 (0.06) |
| 512 | 160 | 400000 | 5000 | 207.57 ( 0.29) | 207.32 | 105.66 ( 0.25) | 105.51 | 2.71 (0.04) |
| 512 | 240 | 200000 | 5000 | 283.98 ( 0.37) | 283.34 | 153.25 ( 0.39) | 152.64 | 2.77 (0.06) |
| 512 | 320 | 200000 | 5000 | 331.65 ( 0.31) | 332.27 | 190.97 ( 0.41) | 191.78 | 2.62 (0.05) |
| 512 | 400 | 200000 | 5000 | 364.43 ( 0.25) | 364.41 | 224.58 ( 0.43) | 224.67 | 2.44 (0.05) |
| 512 | 480 | 200000 | 5000 | 386.93 ( 0.21) | 386.86 | 253.53 ( 0.45) | 253.33 | 2.32 (0.05) |
| 1024 | 80 | 44399 | 5000 | 106.00 ( 0.46) | 106.00 | 55.28 ( 1.39) | 53.00 | 0.91 (0.02) |
| 1024 | 160 | 50000 | 5000 | 213.40 ( 1.01) | 212.62 | 106.98 ( 1.25) | 106.33 | 1.67 (0.06) |
| 1024 | 320 | 400000 | 5000 | 414.81 ( 0.62) | 415.21 | 211.02 ( 0.54) | 211.34 | 3.13 (0.05) |
| 2048 | 640 | 400000 | 5000 | 832.50 ( 1.33) | 830.99 | 424.63 ( 1.15) | 423.00 | 3.58 (0.06) |
| 4096 | 1280 | 400000 | 5000 | 1665.37 ( 2.83) | 1662.55 | 849.15 ( 2.46) | 846.31 | 4.05 (0.07) |
| 8192 | 2560 | 200000 | 5000 | 3308.68 ( 8.45) | 3325.66 | 1678.94 ( 7.34) | 1692.94 | 4.43 (0.12) |
| 16384 | 5120 | 100000 | 5000 | 6614.63 (25.93) | 6651.90 | 3354.28 (22.46) | 3386.19 | 5.09 (0.21) |

Table 1: The results of our runs for the one-dimensional $O(4)$ model. "Sweeps" is the total number of MGMC iterations performed; "Discard" is the number of iterations discarded prior to beginning the analysis. Error bar (one standard deviation) is shown in parentheses. The exact values are obtained from [12].

| $L_{min}$ | $z_{int,\mathcal{M}^2}$ | $A$ | $\chi^2$ | | |
|---|---|---|---|---|---|
| | | | \multicolumn{3}{c}{$d = 1$ $O(4)$ model} |

| $L_{min}$ | $z_{int,\mathcal{M}^2}$ | $A$ | $\chi^2$ | | |
|---|---|---|---|---|---|
| 128 | 0.223 ( 0.005) | 0.6396 ( 0.0219) | 53.264 ( | 6 DF, level = | 0.000 %) |
| 256 | 0.200 ( 0.006) | 0.7647 ( 0.0344) | 15.758 ( | 5 DF, level = | 0.757 %) |
| 512 | 0.184 ( 0.008) | 0.8715 ( 0.0504) | 2.834 ( | 4 DF, level = | 58.602 %) |
| **1024** | **0.174 (0.011)** | **0.9460 (0.0829)** | **1.285 (** | **3 DF, level =** | **73.276 %)** |
| 2048 | 0.163 ( 0.017) | 1.0352 ( 0.1459) | 0.618 ( | 2 DF, level = | 73.431 %) |
| 4096 | 0.155 ( 0.029) | 1.1144 ( 0.2787) | 0.490 ( | 1 DF, level = | 48.378 %) |
| 8192 | 0.200 ( 0.071) | 0.7283 ( 0.4784) | 0.000 ( | 0 DF, level = | 100.000 %) |

Table 2: Weighted least-squares fit for $\tau_{int,\mathcal{M}^2} = A L^{z_{int,\mathcal{M}^2}}$ at $\beta/L = 40/128 = 0.3125$, using lattice sizes $L \geq L_{min}$. Errors in parentheses represent one standard deviation, and "DF" stands for "degrees of freedom". Confidence level is the probability that $\chi^2$ would equal or exceed the observed value, assuming that the underlying statistical model ("null hypothesis") is correct. The line in boldface marks our preferred fit.

| $d = 1$ $O(4)$ model |||| 
|---|---|---|---|
| $L_{min}$ | $A$ | $B$ | $\chi^2$ |
| **128** | **−1.456 (0.090)** | **0.6623 (0.0145)** | **1.678 (6 DF, level = 94.681 %)** |
| 256 | −1.410 ( 0.139) | 0.6560 ( 0.0207) | 1.493 ( 5 DF, level = 91.392 %) |
| 512 | −1.310 ( 0.201) | 0.6428 ( 0.0282) | 1.016 ( 4 DF, level = 90.737 %) |
| 1024 | −1.446 ( 0.336) | 0.6598 ( 0.0438) | 0.759 ( 3 DF, level = 85.931 %) |
| 2048 | −1.512 ( 0.593) | 0.6676 ( 0.0729) | 0.741 ( 2 DF, level = 69.042 %) |
| 4096 | −1.585 ( 1.153) | 0.6759 ( 0.1341) | 0.735 ( 1 DF, level = 39.113 %) |
| 8192 | −4.150 ( 3.205) | 0.9522 ( 0.3489) | 0.000 ( 0 DF, level = 100.000 %) |

Table 3: Weighted least-squares fit for $\tau_{int,\mathcal{M}^2} = A + B \log L$ at $\beta/L = 40/128 = 0.3125$, using lattice sizes $L \geq L_{min}$. Errors in parentheses represent one standard deviation, and "DF" stands for "degrees of freedom". Confidence level is the probability that $\chi^2$ would equal or exceed the observed value, assuming that the underlying statistical model ("null hypothesis") is correct. The line in boldface marks our preferred fit.

| $d = 1$ $O(4)$ model | | | |
|---|---|---|---|
| $L$ | $\tau_{int,\mathcal{M}^2}$ | $\tau_{exp,\mathcal{M}^2}$ | $\tau_{int,\mathcal{M}^2}/\tau_{exp,\mathcal{M}^2}$ |
| 128 | 1.75 (0.03) | 1.9 (0.2) | 0.92 |
| 256 | 2.21 (0.04) | 2.4 (0.3) | 0.92 |
| 512 | 2.71 (0.04) | 3.1 (0.2) | 0.87 |
| 1024 | 3.13 (0.05) | 3.4 (0.2) | 0.92 |
| 2048 | 3.58 (0.06) | 4.0 (0.3) | 0.90 |
| 4096 | 4.05 (0.07) | 4.5 (0.3) | 0.90 |
| 8192 | 4.43 (0.12) | 4.9 (0.4) | 0.90 |
| 16384 | 5.09 (0.21) | 5.8 (0.4) | 0.88 |

Table 4: Estimates of $\tau_{int,\mathcal{M}^2}$ and $\tau_{exp,\mathcal{M}^2}$ for the points with $\beta/L = 0.3125$. See text for the method of calculation. Errors in parentheses represent one standard deviation.

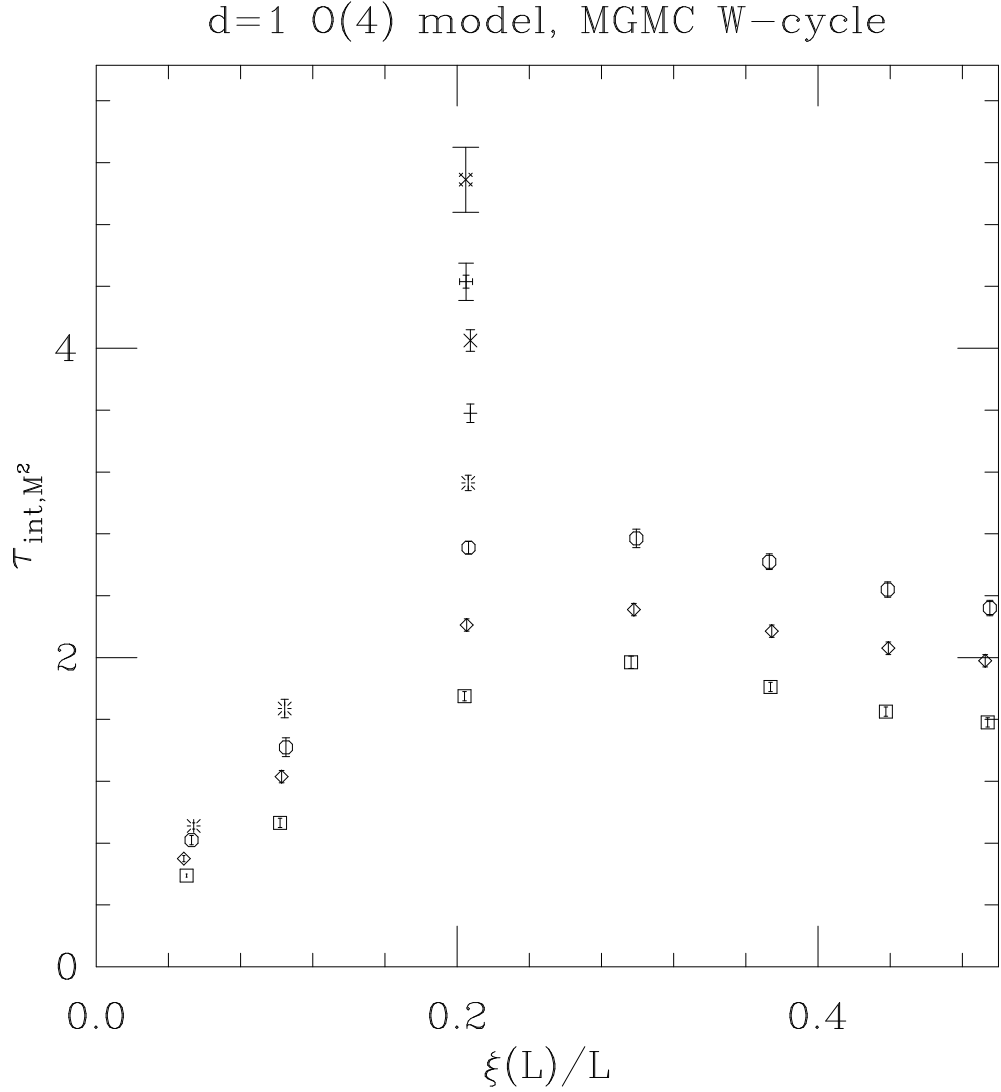

Figure 1: Plot of $\tau_{int,\mathcal{M}^2}$ versus $\xi(L)/L$ for the W-cycle MGMC algorithm, for lattice sizes $L = 128$ ($\square$), $L = 256$ ($\diamondsuit$), $L = 512$ ($\bigcirc$), $L = 1024$ ($*$), $L = 2048$ ($+$), $L = 4096$ ($\times$), $L = 8192$ ($\boxplus$), $L = 16384$ ($\circledast$).

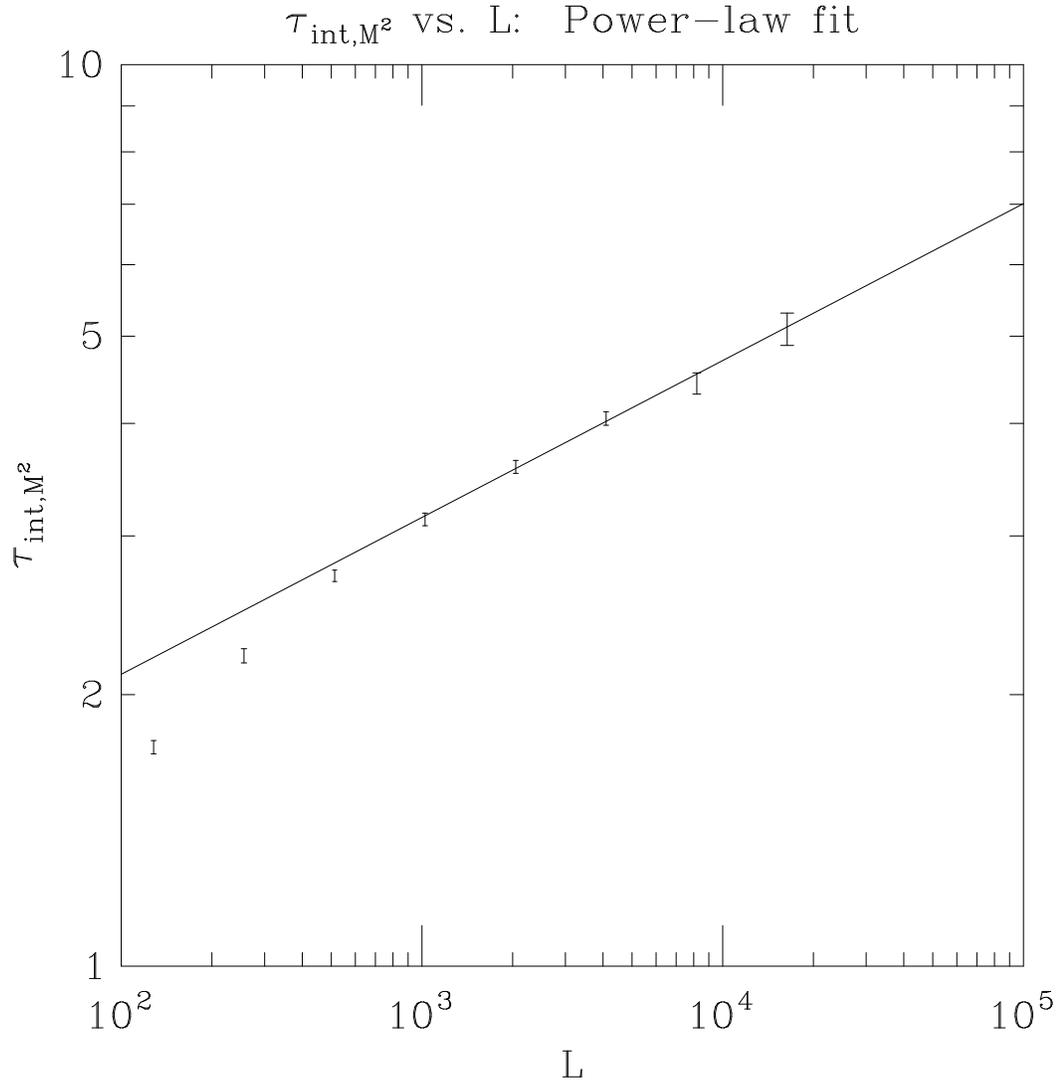

Figure 2: Test of power-law Ansatz $\tau_{int,\mathcal{M}^2} = A\,L^{z_{int,\mathcal{M}^2}}$, for data at $\beta/L = 0.3125$. The straight line is the least-squares fit $\tau_{int,\mathcal{M}^2} = 0.9460\,L^{0.174}$, from Table 2 with $L_{min} = 1024$.

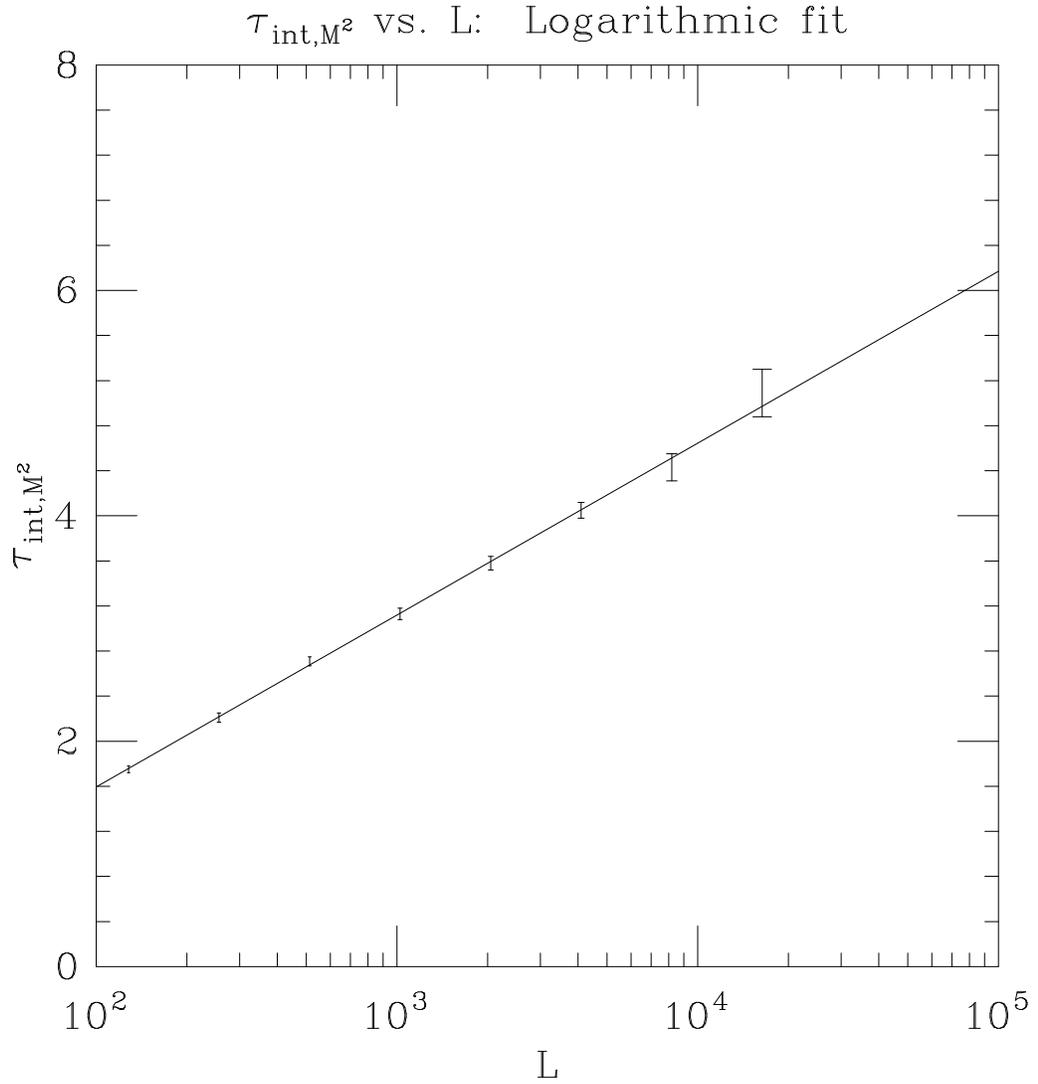

Figure 3: Test of logarithmic Ansatz $\tau_{int,\mathcal{M}^2} = A + B \log L$, for data at $\beta/L = 0.3125$. The straight line is the least-squares fit $\tau_{int,\mathcal{M}^2} = -1.456 + 0.6623 \log L$, from Table 3 with $L_{min} = 128$.

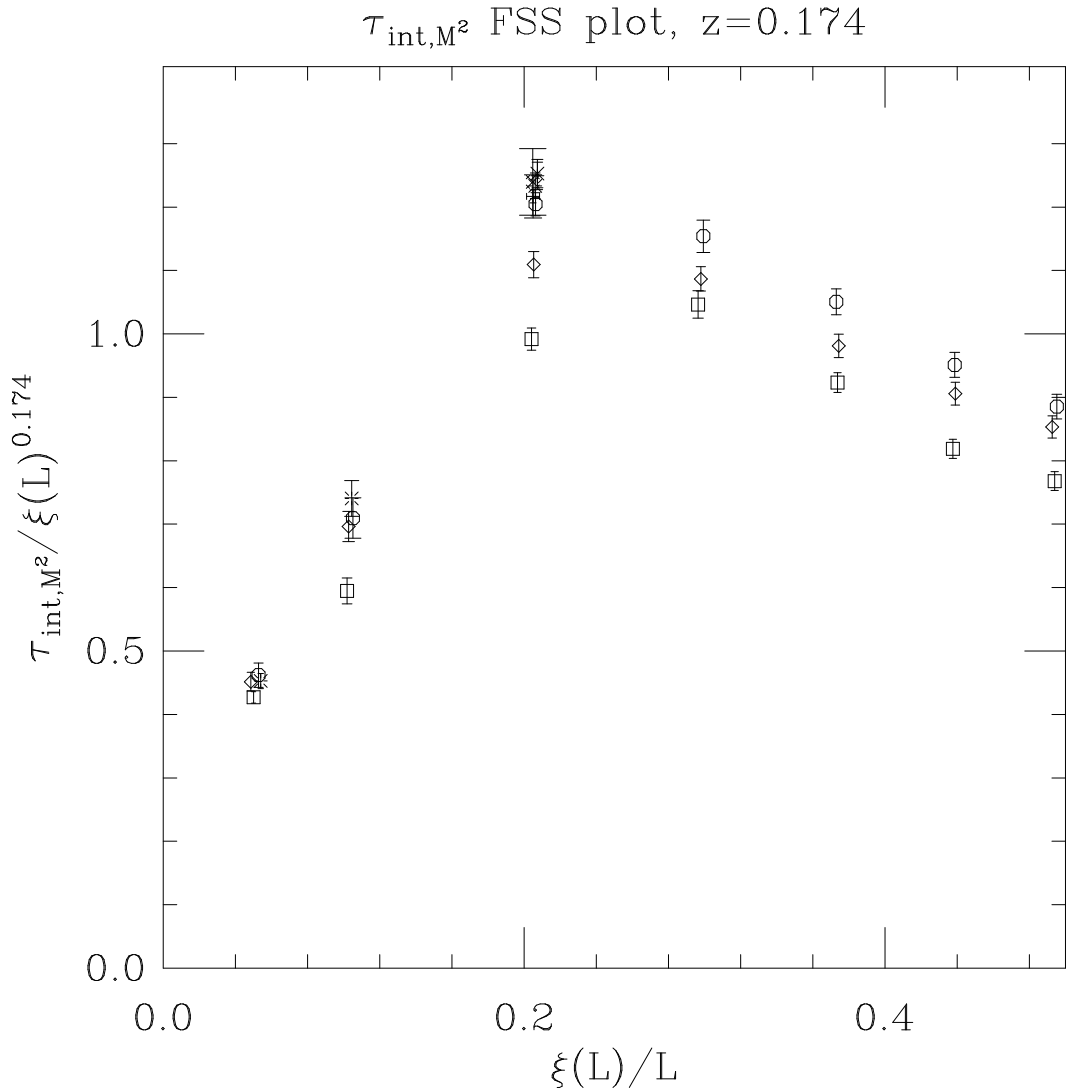

Figure 4: Finite-size-scaling plot of $\tau_{int,\mathcal{M}^2}/\xi^{z_{int,\mathcal{M}^2}}$ versus $\xi(L)/L$ for the W-cycle MGMC algorithm, for lattice sizes $L = 128$ ($\square$), $L = 256$ ($\diamond$), $L = 512$ ($\bigcirc$), $L = 1024$ ($*$), $L = 2048$ ($+$), $L = 4096$ ($\times$), $L = 8192$ ($\boxplus$), $L = 16384$ ($\circledast$). Here $z_{int,\mathcal{M}^2} = 0.174$.

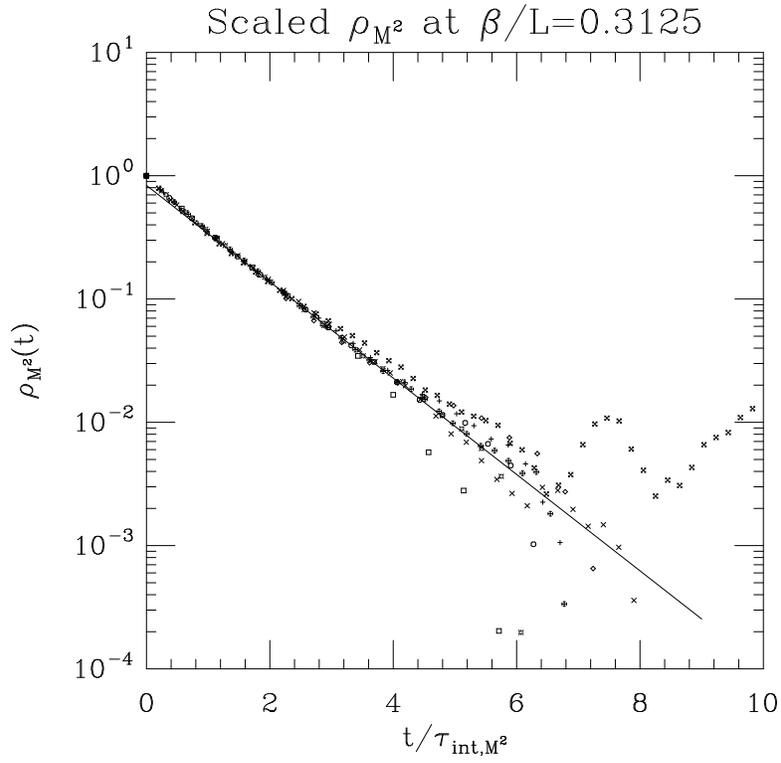

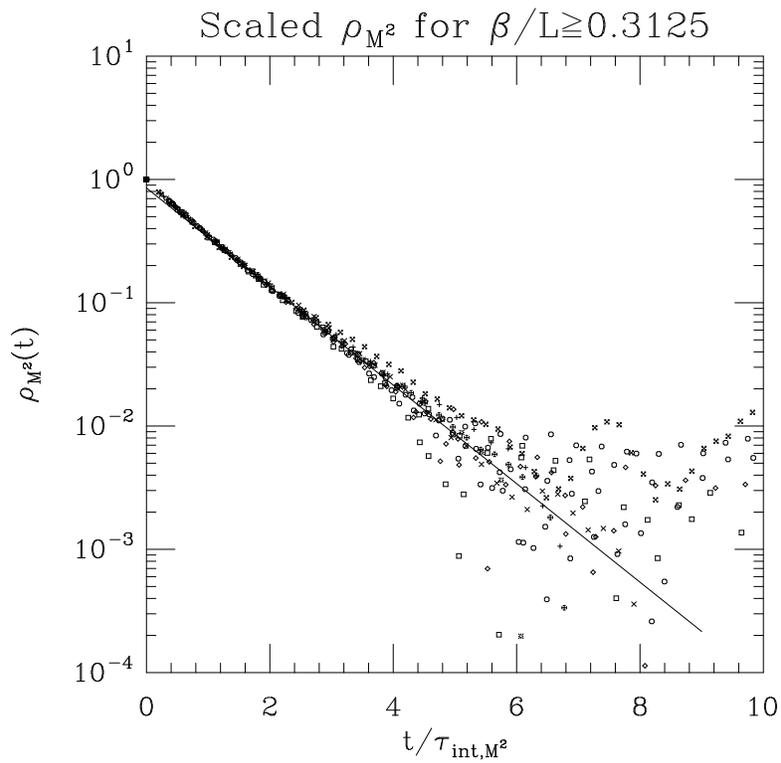

Figure 5: Autocorrelation function $\rho_{M^2}$ as a function of time rescaled by $\tau_{int,\mathcal{M}^2}$, for the case of (a) $\beta/L = 0.3125$, and (b) all points with $\beta/L \geq 0.3125$. The straight lines are linear fits to the interval $1.5 \leq t/\tau_{int,\mathcal{M}^2} \leq 3$.